\documentclass[aps,prb,preprint]{revtex4}
\usepackage{graphicx}
\usepackage{amsmath}
\usepackage{scalerel,stackengine}
\stackMath
\newcommand\reallywidehat[1]{%
\savestack{\tmpbox}{\stretchto{%
  \scaleto{%
    \scalerel*[\widthof{\ensuremath{#1}}]{\kern-.6pt\bigwedge\kern-.6pt}%
    {\rule[-\textheight/2]{1ex}{\textheight}}
  }{\textheight}%
}{0.5ex}}%
\stackon[1pt]{#1}{\tmpbox}%
}

\begin{document}

\title{On the Dirac Quantisation rules and the trace anomaly}

\author{T. C. Choy}
\email{tuckvk3cca@gmail.com}
\affiliation{Samy Maroun Center for Space, Time and the Quantum, Parc Maraveyre B${\hat a}$t. 1, 13260, Cassis, Bouches du Rh${\hat o}$ne, France  and \\
Departmento de Fisica-CIOyN, Universidad de Murcia, Murcia 30071, Spain.}

\begin{abstract}
In this article I shall clarify various aspects of the Dirac quantisation rules of 1930\cite{Dirac}, namely (i) the choice of antisymmetric Poisson brackets, (ii) the first quantisation Rule 1 (iii) the second quantisation Rule 2, and their relations to the trace anomaly. In fact in 1925 Dirac already had a preliminarily formulation of these rules \cite{Dirac3}. Using them, he had independently rediscovered the Born-Jordan quantisation rule \cite{BornJordan1925} and called it the quantum condition.  This is the best known and undoubtedly most significant of the canonical quantisation rules of quantum mechanics. We shall discuss several violations of the Poisson-Lie algebra (assumed by Dirac), starting from antisymmetry, which is the first criterion for defining a Lie algebra. Similar violations also occur for the Leibniz's rule and the Jacobi identity, the latter we shall also prove for all our quantum Poisson brackets.  That none of these violations jeopardised Dirac's ingenious original derivation \cite{Dirac} of his first quantisation Rule 1, is quite remarkable. This is because the violations are all of higher orders in $\hbar$.    We shall further show that (ii) does not automatically lead to a trace anomaly for certain bounded integrable operators. Several issues that are both pedagogical and foundational arising from this study show that quantum mechanics is still not a finished product. I shall briefly mention some attempts and options to complete its development.

\end{abstract}

\maketitle

\section{Introduction}
\label{Introduction}

A recent column by  Mahajan \cite{Mahajan}  in this journal concerning the quantum trace anomaly, prompted me to pen this article which
addresses several issues that are perhaps not widely known or even discussed in textbooks. These issues concern Dirac quantisation, its subtleties and the trace anomaly. The quantum trace anomaly was first discovered by Born and Jordan in 1925\cite{BornJordan1925}. It arises from considering the trace of the Dirac Quantum Condition \cite{Dirac} or the Born-Jordan canonical commutation relation (CCR) \cite{BornJordan1925}:
\begin{equation}
\ [{\hat q},{\hat p}\ ]= i\hbar\ {\hat I} \ .
\label{BJ Rule}
\end{equation}
The LHS of this equation\cite{endnote1} is the commutator of the two canonically conjugate quantum operators ${\hat q}$ and ${\hat p}$  for coordinate and momentum, and the RHS contains the unit operator ${\hat I}$. Eqn(\ref{BJ Rule}) was first given by Born and Jordan as a physical law in their seminal paper \cite{BornJordan1925} which was a development from the earlier ideas of Heisenberg \cite{Heisenberg}. Both these works were in some respects semi-empirical and were mathematical extensions of the old quantum rules of Wilson \cite{Wilson1915} and Sommerfeld \cite{Sommerfeld1915}. As is well known, Heisenberg \cite{Heisenberg} was at that time ``fabricating" the new mechanics, by introducing some key ideas, guided notably by Ritz's combination principle of atomic spectral data \cite{Aitchinson2004}.  It was Dirac who made the first major attempt at a formulation of quantum mechanics from fundamental theoretical principles \cite{Dirac}.

Over the years, I have observed that the trace anomaly also signifies another aberration. Never in the history of physics it seems, has such a profound physical law as eqn(\ref{BJ Rule}) been formulated that is mathematically flawed, but still has such an astounding physical impact and technological significance.  Yet this anomaly is so grossly ignored in textbooks, historical volumes \cite{MehraRechenberg1982} and journal articles \cite{Berstein2005,Fedak2009}. Even lately, it has received a mere one paragraph mention \cite{Mahajan} after almost one hundred years of quantum mechanics history.  This article is my personal attempt towards a remedy. It should be read as a primer on the subject of Dirac quantisation, for those who have not examined the issues here in depth or for those just curious about historical and modern developments in quantum theory. No attempts will be made to maintain pure mathematical rigour; in particular convergence issues. For this the reader should consult other references \cite{Emch1972,Hall2013}.

\section{What is the trace anomaly?}
\label{whatisTA}
Briefly, we shall start with the naive argument. Taking the trace on the LHS of eqn(\ref{BJ Rule}) will give us zero because of the cyclic invariance of the trace in matrix algebra as originally developed by Born and Jordan \cite{BornJordan1925}.  This is not so for the RHS hence there is a contradiction. Born and Jordan in their seminal paper \cite{BornJordan1925} stated in the text and a footnote that the trace anomaly implies that the quantum operators must be {\it both} (my italics)infinite dimensional matrices and also unbounded.  The arguments presented by Mahajan \cite{Mahajan} is that the matrices must be infinite dimensional, echoing the words of Born and Jordan, and that somehow this will fix the problem, without mentioning unboundedness. We shall see later that both these assumptions will not fix the problem. For some unknown reason, Born and Jordan's statement was taken by some in subsequent generations to imply that bounded operators are therefore forbidden in quantum theory.  Neither they nor subsequent writers ever go into depth as to how the anomaly is resolved by considering infinite dimensional Hilbert spaces of unbounded operators. Dirac unfortunately was silent about the subject.  We might perhaps speculate as to why at the end of this article.

Some mathematicians following the axiomatic foundations of von Neumann \cite{vonNeumann1927} and algebraic generalisations see this difficulty as best overcome by treating only systems with an infinite degree of freedom, abandoning Fock Space and adopting a more sophisticated C* algebra formalism based on the commuting (CCR) or  anticommuting (ACR) rules of field theory; see for example Emch(1972) \cite{Emch1972}. These do not however offer much insight into the resolution of the trace anomaly. Therefore let us first examine this anomaly issue more closely.  Taking the trace of eqn(\ref{BJ Rule}) we have for the LHS:
\begin{eqnarray}
Tr \ [{\hat q},{\hat p}\ ] &=&  \int dq\ dq'\ <q'|[{\hat q},{\hat p}\ ]|q>\delta(q-q')\  ,  \nonumber \\
    & = &\int dq\ dq'\ (q'-q) <q'|{\hat p} |q> \delta(q-q').
\label{BJ Rule Trace}
\end{eqnarray}
 The second line in eqn(\ref{BJ Rule Trace}) follows because Hermitian operators can act either from the left or from the right, giving their respective {\it real} eigenvalues. If the matrix elements $<q'|{\hat p} |q>$ are all bounded and well behaved, then the LHS is zero. However this is not the case here and to proceed further, let us introduce a complete set of momentum states $|p>$ with the wave-functions \cite{Dirac} $<q|p>=\frac{1}{\sqrt{\hbar}} e^{i\frac{pq}{\hbar}}$. Then:
\begin{eqnarray}
Tr \ [{\hat q},{\hat p}\ ] & =& \int dq\ dq' dp \ (q'-q)p  <q'|p><p|q> \delta(q-q'), \nonumber \\
& =& \int dq\ dq' \frac{dp}{\hbar} \ (q'-q) p\  e^{\frac{i(q'-q)p}{\hbar}} \delta(q-q').
\label{BJ Rule TraceA}
\end{eqnarray}
It is straightforward to see that if the $q$ or $q'$ integration is done first, the result is zero.  On the other hand if the $p$ integration is done first, then the result is divergent.  In other words, the trace of the left hand side of eqn(\ref{BJ Rule}) is ill-defined.
Now for the RHS we have:
\begin{eqnarray}
i\hbar \ Tr \  {\hat I}\  &=& i\hbar  \int dq\ dq'\ <q'|{\hat I}  ]|q>\delta(q-q')\  , \nonumber \\
    & = & i\hbar \int dq\ dq'\ <q'|q> \delta(q-q').
\label{BJ Rule Trace2}
\end{eqnarray}
This integral is infinite since $<q|q'>=\delta(q-q')$ due to normalisation. As one can see, in the Hilbert space in which we have defined all our operators, the trace on the RHS of eqn(\ref{BJ Rule}) i.e.  eqn(\ref{BJ Rule Trace2}) is infinite while that on the LHS of eqn(\ref{BJ Rule}) i.e. eqn(\ref{BJ Rule TraceA}) is ill-defined.  The suggestion that infinite dimensional matrices consisting of bounded matrices \cite{Mahajan} or unbounded ones \cite{BornJordan1925} will somehow resolve this anomaly is untenable.
None of the operators in eqn(\ref{BJ Rule}) belong to the trace class $Tr \sqrt{A^\dag A}<\infty$ which is at the heart of the problem of the trace anomaly.

However in condensed matter physics, we often deal with bounded operators with or without cyclic (i.e. Born von-Karman) boundary conditions. This is particularly important because Dirac's quantisation Rule 1 (see eqn(\ref{Dirac rule1}) later), can be used to derive the correct angular momentum operator commutation relations. These are in fact {\it without} trace anomalies, as demonstrated by Dirac himself \cite{Dirac2}.  Other authors such as Costella (1995) \cite{Costella1995} suggested that the trace anomaly could be resolved by discretization to a lattice which in the Schrodinger representation implies that the RHS is actually off-diagonal. Unfortunately such a scheme violates the essential theorem that in the energy representation the canonical commutator eqn(\ref{BJ Rule}) is a constant of motion and must be strictly diagonal, Born and Jordan 1925 \cite{BornJordan1925}. In this paper we shall first show that Dirac made a specific choice \cite{Dirac} in his definition of the quantum Poisson brackets. This choice can be modified by antisymmetrisation but does reveal that several different definitions of quantum Poisson brackets can be constructed.  Then we shall return to the trace anomaly and later also other issues, concluding with some discussions.

\section{Poisson-Lie algebra}
\label{Possion-Lie}
As is well known, Dirac \cite{Dirac} assumed that his quantum Poisson brackets obey the same Lie algebra properties as in classical mechanics.  These properties are given by:
\begin{eqnarray}
\{u,v\} &=& -\{v,u\} \ \label{antisymmetry} \\
\{u,v{_1}+v{_2}\} &=& \{u,v{_1}\}+\{u,v{_2}\} \ \label{biliearity1}  \\
\{u{_1}+u{_2,v}\} &=& \{u{_1},v\}+\{u{_2},v\} \ \label{biliearity2} \ \\
\{uv,w\} &=& u\{v,w\} + \{u,w\}v \ \label{LeibnizRule} \\
\{\{u,v\},w\} &+& \{\{v,w\},u\}  + \{\{w,u\},v\}  =0\ \label{JacobiIdentity}
\end{eqnarray}
The first of these eqn(\ref{antisymmetry}) is the antisymmetry criterion, while eqn(\ref{biliearity1}) and eqn(\ref{biliearity2}) are the bi-linearity criteria. Note that antisymmetry and bi-linearity are closely related. If we let $w=u+v$ and $\{w,w\}=\{u,u\}=\{v,v\}=0$, it will follow from bi-linearity that $\{u,v\}=-\{v,u\}$. The fourth eqn(\ref{LeibnizRule}) is Leibniz's rule or the derivative property and the last eqn(\ref{JacobiIdentity}) is the Jacobi identity, (valid for any cyclic permutations of the three quantities $u,v,w$ ). Both these properties are easy to prove for a commutator bracket matrix representation of the algebra, but the last property in particular was notoriously difficult to prove in classical mechanics \cite{LandauLifshitz1}.  In fact it was not until the year  2000 that an elegant simplified proof was first published in this very journal \cite{Lemos2000}, see Appendix 1. There it will be shown that antisymmetry is fundamental to proving the Jacobi identity for all our Poisson brackets in this paper.  These algebraic properties uniquely define a Poisson-Lie algebra \cite{endnote2}. In classical mechanics the algebra guarantees the invariance of Poisson brackets under {\it all} canonical transformations thereby preserving the Hamilton equations of motion, presumed to be true also upon quantisation (see later).  Dirac appreciated this very early on as a profound statement of Heisenberg's central hypothesis \cite{Heisenberg} for the advent of quantum mechanics. In particular by exploiting Leibniz's rule for a bracket containing four operators he arrived at his first famous quantisation rule \cite{Dirac}, which we shall refer to as Dirac's Rule 1 \cite{endnote3}:
\begin{equation}
[{\hat u},{\hat v}]= i\hbar \{{\hat u},{\hat v}\}_{P_1}\ ,
\label{Dirac rule1}
\end{equation}
where the LHS is the usual commutator bracket of two quantities which are functions of canonically conjugate dynamic variables and the RHS is the quantum Poisson bracket of the same quantities, with  $\hbar $ the Planck's constant. Dirac chose to define this operator as the quantum operator \cite{Dirac}:
\begin{equation}
\{{\hat u },{\hat v}\}_{P_1}=\bigl ({\partial \hat u \over \partial \hat q} {\partial \hat v \over \partial \hat p}  - {\partial \hat u \over \partial \hat p} {\partial \hat v \over \partial \hat q}\bigr )\ . \\
\label{P1definition}
\end{equation}
In operator form \cite{diffOperator}, eqn(\ref{Dirac rule1}) and eqn(\ref{P1definition}) have a contradiction.  Closer examination shows that the quantum operator form, in eqn(\ref{P1definition}) is neither symmetric or antisymmetric.  Therefore Dirac's assumption that his quantum bracket $P1$ satisfies a Poisson-Lie algebra is strictly incorrect.  This raises two questions.  (i) Can we use other (explicitly antisymmetric) choices for the quantum Poisson brackets (see for example Lanczos \cite{Lanczos}) such as:
\begin{equation}
\{{\hat u },{\hat v}\}_{P_2}=\bigl ({\partial \hat u \over \partial \hat q} {\partial \hat v \over \partial \hat p} - {\partial \hat v \over \partial \hat q} {\partial \hat u \over \partial \hat p}\bigr )\ , \\
\label{P2definition}
\end{equation}
or:
\begin{equation}
\{{\hat u },{\hat v}\}_{P_3}=\bigl ({\partial \hat u \over \partial \hat p} {\partial \hat v \over \partial \hat q} - {\partial \hat v \over \partial \hat p} {\partial \hat u \over \partial \hat q}\bigr )\ , \\
\label{P3definition}
\end{equation}
which {\it a priori} do not have to be equivalent, see for example Shewell \cite{Shewell1959}. The second question is should we retain Dirac's definition and antisymmetrise it so that Dirac's Rule 1, eqn(\ref{Dirac rule1}) is no longer contradictory and therefore satisfies the first criterion of the Poisson-Lie algebra eqn(\ref{antisymmetry})?  Let us look more closely at Dirac's quantum Poisson bracket.  It is in fact straightforward to show that while strictly non-antisymmetric, the violating terms by his Rule 1 are of the first order in $\hbar$:
\begin{eqnarray}
\{{\hat v},{\hat u}\}_{P_1} &=& -\{{\hat v},{\hat u}\}_{P_1} + [\frac{\partial {\hat v}}{\partial {\hat q}},\frac{\partial {\hat u}}{\partial {\hat p}}]-[\frac{\partial {\hat v}}{\partial {\hat p}},\frac{\partial {\hat u}}{\partial {\hat q}}]\ \nonumber \\
 &=& -\{{\hat v},{\hat u}\}_{P_1} + i\hbar \{\frac{\partial {\hat v}}{\partial {\hat q}},\frac{\partial {\hat u}}{\partial {\hat p}}\}_{P_1}-i\hbar \{\frac{\partial {\hat v}}{\partial {\hat p}},\frac{\partial {\hat u}}{\partial {\hat q}}\}_{P_1} .
\label{P1other terms}
\end{eqnarray}
Here we have replaced the commutators with Poisson brackets in accordance with Rule 1. Substituting the second of eqn(\ref{P1other terms}) into eqn(\ref{Dirac rule1}), we can now see that the violations to antisymmetry there are of order $\hbar^2$. In fact for simple functions $\hat u=\hat q$ and $\hat v=\hat p$ or vice-versa, evaluating the brackets in eqn(\ref{P1other terms}) shows that these do not lead to any violations. Hence these two objects belong to a sub-algebra that is antisymmetric. In a later section we shall also look at violations of the other criteria for a Lie algebra which are also of order $\hbar^2$. In the meantime the reader is reminded that the trace anomaly is an anomaly of order $\hbar$.

\section{Antisymmetrised Poisson Bracket}
\label{section II}

As we have seen, Dirac's definition of the quantum Poisson bracket eqn(\ref{Dirac rule1}), in fact has no particular symmetry or antisymmetry.  As Dirac was well known to be a very particular and pedantic person, and given that he has chosen to define his Poisson brackets specifically in many places consistently \cite{DiracLectures} there must be good reasons for this.  Unfortunately I could find no sources in which he explicitly spelled out the reason for his choice, so we are left to speculate. It is very likely that he must have obtained it from the following argument.  For any phase space function operator ${\hat f}$, his choice of quantum Poisson brackets could be obtained from:
\begin{equation}
{d {\hat f}\over dt}= {\partial {\hat f}\over \partial t} + {\partial \hat f \over \partial \hat q}{\dot {\hat q}} +  {\partial \hat f \over \partial \hat p}{ \dot {\hat p}}\ , \\
\end{equation}
maintaining operator orderings.

Hence from Hamilton's equations:
\begin{equation}
{\dot {\hat q}} =  {\partial \hat H \over \partial \hat p}\ ,  \ {\dot {\hat p}} =  -{\partial \hat H \over \partial \hat q} ,\\
\label{Hamilton1}
\end{equation}
this gives:
\begin{equation}
{d {\hat f}\over dt}= {\partial {\hat f}\over \partial t} + \{{\hat f },{\hat H}\}_{P_1} \ , \\
\label{Hamilton2a}
\end{equation}
and therefore:
\begin{equation}
\{{\hat f },{\hat H}\}_{P_1}=\bigl({\partial \hat f \over \partial \hat q} {\partial \hat H \over \partial \hat p}  - {\partial \hat f \over \partial \hat p} {\partial \hat H \over \partial \hat q} \bigr)\ , \\
\label{PoissonBracketP1}
\end{equation}
which as the reader can see has {\it no} particular symmetry properties, being neither symmetric nor antisymmetric in $\hat f, \hat H $.  From this Dirac must have generalized to the generic Poisson bracket $P_1$, i.e. eqn(\ref{P1definition}), by replacing ${\hat H}$ with any arbitrary generator ${\hat g}$ of canonical transformations. However it seems logical that one could antisymmetrise this operator for quantisation so that Dirac's first quantisation rule should now read:
\begin{equation}
[{\hat u},{\hat v}]=i\hbar \{{\hat u},{\hat v}\}_{P_1 Antisymm}\ .
\label{Dirac rule2}
\end{equation}
The reader can confirm by straightforward manipulations that this antisymmetrised operator is  now in fact given by:
\begin{equation}
\{{\hat u},{\hat v}\}_{P_1 Antisymm}= \frac{1}{2}\Bigl[\{{\hat u},{\hat v}\}_{P_2}+\{{\hat v},{\hat u}\}_{P_3}\Bigr]\ .
\label{PoissonBracketP5}
\end{equation}
Eqn(\ref{Dirac rule2}) and eqn(\ref{PoissonBracketP5}) constitute a modification to Dirac's Quantisation Rule 1 eqn(\ref{Dirac rule1}). As we have seen, for antisymmetry the difference between $P_1$ and ${P_{1 Antisymm}}$ are of a higher order in $\hbar$, so laboratory experimental observations will be a great challenge. Since we must focus here on Dirac's theory, unless specifically stated, we shall pay no further attention to quantum bracket types $P_2$,$P_3$ or indeed ${P_{1 Antisymm}}$ and ignore our bracket subscripts from now on. They all in fact represent alternative theories that will take us too far afield to discuss. Suffice to say, for example using bracket type $P_3$ for quantisation will require a modification of Rule 1 with the minus sign \cite{endnote3} to be empirically correct. Therefore from here on, all Poisson brackets are understood to be of type $P_1$, and we shall drop all subscripts unless explicitly required.

The problem now reduces to one of evaluating the quantum Poisson brackets defined in the last section. Here Dirac made a further (simplifying) assumption \cite{Dirac,Dirac3} that the canonical coordinate and momentum quantum Poisson brackets are just identical ``in value" to the classical ones \cite{Dirac}. This is not quite correct. In fact in the one dimensional case here he basically proposed Rule 2:
\begin{equation}
  \{{\hat u},{\hat v}\}=\reallywidehat{{\{u,v\}}}\
\label{Dirac Quantum Condition}
\end{equation}
where the large hat indicates an operator. That is, first evaluate the classical Poisson bracket then turn the result into an operator. We shall see later that for the type of functions we shall consider, this rule is quite unnecessary. For ${\hat u}={\hat q},{\hat v}={\hat p}$  we will get a unit operator which gives the Born-Jordan quantisation rule \cite{BornJordan1925} or Dirac quantum condition using {\it either} Rule 1 or Rule 2. which will lead to a trace anomaly. Further there are other inconsistencies discovered in later years from Rule 2 that are by now well known \cite{Shewell1959,Groenewold1946}. The reason for the introduction of Rule 2 by Dirac was rather unclear. In fact up until 1930, he actually defined the quantum Poisson bracket using the commutator relation Rule 1 eqn(\ref{Dirac rule1}) in the opposite direction i.e. from right to left.  Given that the matrix differential calculus was already set up by Born and Jordan in 1925 \cite{BornJordan1925}, Rule 2 seems redundant for the quantum Poisson brackets can be evaluated directly with this calculus. The proposal of this rule had led to many misunderstandings and inconsistencies dating back to Groenewold \cite{Groenewold1946, Kauffman2011}.  My speculation is that Dirac invented this rule to avoid the tedious problem of symmetrisation, which he never considered: see later.

\section{Trace anomaly for the Dirac quantisation Rule 1}
\label{section III}

As this stage it seems obvious that one should re-examine the trace anomaly in the context of the first Dirac quantisation rule eqn(\ref{Dirac rule1}) for trace class operators \cite{traceclassI}.  To see how this works, let us simplify matters by first considering conjugate operators ${\hat u}({\hat q})$ and ${\hat v}({\hat p})$ only, i.e. the first is only a function of $q$ and the second a function of $p$. Then, following eqn(\ref{BJ Rule Trace}) we now have:
\begin{eqnarray}
Tr \ [{\hat u},{\hat v}\ ] &=&  \int dq\ dq'\ <q'|[{\hat u},{\hat v}\ ]|q>\delta(q-q')\  ,  \nonumber \\
    & = &\int dq\ dq'\ (u(q')-u(q)) <q'|{\hat v}({\hat p}) |q> \delta(q-q') ,  \nonumber \\
    & = &\int dq\ dq'\ (u(q')-u(q)) V(q'-q) \delta(q-q').
\label{Dirac Rule Trace1}
\end{eqnarray}
Here we have introduced a complete set of momentum states, and as before, the matrix element:
\begin{equation}
<q'|{\hat v}({\hat p})|q>=V(q'-q)=\int \frac{dp}{\hbar}\ v(p)\ e^{\frac{i(q'-q)p}{\hbar}}
\label{VFourierTransform}
\end{equation}
is consequently a Fourier transform. We shall assume that it exists and is finite; a sufficient condition is that $v(p)$ is bounded and integrable.  Eqn(\ref{Dirac Rule Trace1}) is then well defined and is identically zero. The trace on the RHS of eqn(\ref{Dirac rule1}) is:
\begin{equation}
i\hbar\ Tr \ \{{\hat u },{\hat v}\}\  =  i \hbar \int dq\ dq'\ <q'|{\partial \hat u \over \partial \hat q} {\partial \hat v \over \partial \hat p}|q>\delta(q-q').
\label{Dirac Rule Trace2a}
\end{equation}
Now for the case of ${\hat u }={\hat q }$ and ${\hat v }={\hat p }$, eqn(\ref{Dirac Rule Trace1}) and eqn(\ref{Dirac Rule Trace2a}) will return to eqn(\ref{BJ Rule TraceA}) and eqn(\ref{BJ Rule Trace2}) respectively as before. So we can conclude that Rule 1 is {\it not} free from the trace anomaly for such unbounded operators. However for other arbitrary functions we have instead \cite{diffOperator}:
\begin{eqnarray}
i\hbar\ Tr \ \{{\hat u },{\hat v}\}\ & = &i \hbar \int dq\ dq'\ {\partial  u \over \partial  q'} <q'| {\partial \hat v \over \partial \hat p}    |q> \delta(q-q'), \nonumber \\
    & = &i \hbar \int dq\ dq' dp\ {\partial  u \over \partial  q'} <q'| {\partial \hat v \over \partial \hat p} |p><p|q> \delta(q-q').
\label{Dirac Rule Trace2}
\end{eqnarray}
Once again we have introduced a complete set of states $|p>$, so that:
\begin{eqnarray}
i\hbar\ Tr \ \{{\hat u },{\hat v}\}\  & = &i \hbar\int dq\ dq' \frac{dp}{\hbar}\ {\partial  u \over \partial  q'} {\partial  v \over \partial  p}\ e^{i\frac{(q'-q)p}{\hbar}}\ \delta(q-q'), \nonumber \\
& = &i\int dq\ ( \ {\partial  u \over \partial  q})\ \int dp\ ({\partial  v \over \partial  p})
\label{Dirac Rule Trace2b}
\end{eqnarray}
as the integration over $q'$ is now trivial and the integrations can be done in any order. As we can see, {\it as long as the second integral is well defined}, then for bounded periodic functions in a box $L$ (e.g. Born von-Karman boundary conditions),  the first integral (which is trivial, being $u |_{-L}^{L} $), makes the whole expression vanish. Note that periodicity need not be a consequence of boundary conditions, but can also follow from the choice of the dynamical variables, such as $u=\sin (q/L), v=\cos (p/{p_F})$ etc, see also Section \ref{Weyl Quantisation}. The vanishing of this first integral is also true for integrable functions that vanish at infinity i.e. $u(\pm L) \rightarrow 0$ as $L \rightarrow \infty$, so there is no trace anomaly for these cases.  The assumption that quantum mechanics must be formulated only in terms of unbounded operators in an infinite dimensional Hilbert space to avoid the trace anomaly is unnecessary and incorrect in the case of Rule 1 \cite{vonNeumannUniqueness}. As long as the operators $u(q)$ and $v(p)$ are of the trace class in eqn(\ref{Dirac rule1}), then Rule 1 is self consistent and we have no trace anomaly.  The reader can be easily convinced of a similar conclusion in the opposite case where now the conjugate operators are ${\hat u}({\hat p})$ and ${\hat v}({\hat q})$ instead.

We can now extend this study further to look at arbitrary (non-separable) functions $u(p,q)$ and $v(p,q)$ which are phase space functions that can be expanded via a Taylor-Maclaurin series of the general type \cite{endnote4}:
\begin{equation}
a(p,q) = \sum_{s,r=0}^{\infty} C_{s,r} p^s q^r,
\label{Taylor series}
\end{equation}
where $ C_{r,s}$ are their respective {\it real} derivative coefficients, which we shall assume to converge and exist. This does not mean however that we can readily transfer this expansion to their associated quantum operators: ${\hat u}({\hat p},{\hat q})$ and ${\hat v}({\hat p},{\hat q})$. As noted earlier the von Neumann prescription does not provide for uniqueness and there are different ways of representing higher power operators such as for example ${\hat p}^2{\hat q}^2$ which under the Dirac quantisation Rule 2 can have inconsistent forms \cite{Shewell1959,Groenewold1946}. Physically one normally seeks a quantum operator from a known classical observable with the form eqn(\ref{Taylor series}), but due to non-commutativity, this can have many terms when written out in full in terms of operators, such as ${\hat p}^3{\hat q}^2$ or ${\hat p}{\hat q}{\hat p}{\hat q}{\hat p}$. However such operators are unacceptable for the following reason. The quantum operator functions $\hat u({\hat p},{\hat q})$ or $\hat v({\hat p},{\hat q})$  must be consistent with Dirac's Rule 1 in the first place or there will be a contradiction. Since functions of the type $\hat u({\hat p},{\hat q})$ are to be arbitrary generators of canonical transformations, a necessary and sufficient condition is that they must be Hermitian {\it and} satisfy the Hamilton-like (consistency) equations of eqn(\ref{Hamilton1}) (cf Appendix 1) i.e.
\begin{eqnarray}
\frac{d {\hat q}}{d \tau } &=&  \{{\hat q},{\hat u} \} =  {\partial \hat u \over \partial \hat p} =\frac{1}{i\hbar} [{\hat q},{\hat u} ], \nonumber \\
\frac{d {\hat p}}{d \tau } &=&  \{{\hat p},{\hat u} \} = - {\partial \hat u \over \partial \hat q} = \frac{1}{i\hbar} [{\hat p},{\hat u} ].
\label{Hamilton2}
\end{eqnarray}
We shall describe this type of function as belonging to the BJ Class, for Born-Jordan or W Class for Weyl \cite{Weyl1931} (see Section \ref{Weyl Quantisation}). (There is a subtle difference in defining the derivatives for these two classes which we shall need to go into later on, see Appendix 2)

Born and Jordan in their seminal paper originally proved their remarkable result eqn(\ref{BJsymmetryrule1}) below using the canonical quantisation rule CCR  eqn(\ref{BJ Rule}) and its generalisations \cite{BornJordan1925}. In order to be consistent with eqn(\ref{Hamilton2}), they proposed that the quantised ${\hat p}^r{\hat q}^s$ operator in eqn(\ref{Taylor series}) must be symmetrised ( sometimes also referred to as ``quantised" in the literature \cite{Gosson2014}) into either one of the two equivalent forms:
\begin{equation}
  {\hat p}^s\ {\hat q}^r \rightarrow \frac{1}{s+1}\sum_{l=0}^{s}{\hat p}^{s-l}\ {\hat q}^r\ {\hat p}^l.
\label{BJsymmetryrule1}
\end{equation}
{\it or}
\begin{equation}
 {\hat p}^s\ {\hat q}^r \rightarrow  \frac{1}{r+1}\sum_{j=0}^{r} {\hat q}^{r-j}\ {\hat p}^s\ {\hat q}^j.
\label{BJsymmetryrule1a}
\end{equation}

In Appendix 2, we shall prove that their proposal remains valid by using Rule 1 alone and is not restricted to the CCR eqn(\ref{BJ Rule}). There we shall also prove that for the Weyl's symmetrisation scheme a similar result follows (see Section \ref{Weyl Quantisation} later).  Weyl has made it quite explicit in his lecture notes \cite{Weyl1931} that eqn(\ref{Hamilton2}) was an essential criterion for his quantisation scheme, but how it operates needs clarification.  As we shall see in Appendix 2, the BJ scheme satisfies eqn(\ref{Hamilton2}) for two types of derivatives, namely the Born-Jordan \cite{BornJordan1925} and Heisenberg \cite{Born-Jordan-Heisenberg1926} derivatives which are equivalent only under the BJ scheme \cite{Gosson2014}.  However, the Weyl's scheme only satisfies eqn(\ref{Hamilton2}) for the Heisenberg \cite{Born-Jordan-Heisenberg1926} derivative, and the two types of derivative are inequivalent under his scheme.  We will not be able to go into these subtle issues in depth in this article. Hence from now on we shall only be considering operator functions of the BJ Class or W Class. These are obtained by substituting either expression eqn(\ref{BJsymmetryrule1}) or eqn(\ref{BJsymmetryrule1a}) and eqn(\ref{Weylsymmetryrule1a}) or eqn(\ref{Weylsymmetryrule1}) (whichever is appropriate and convenient) into eqn(\ref{Taylor series}).

Moreover the algebra of these two classes of functions (with the appropriate derivatives) satisfy the important required property that the product of two such functions or the product of two derivatives belong to the same class, see Appendix 2. Hence each term in the Taylor expansion of these products is also a linear sum of the generic form ${\hat p}^r{\hat q}^s{\hat p}^t$ or ${\hat q}^r{\hat p}^s{\hat q}^t$ as required by theorem I below; for proof see Appendix 2.

The following Theorem I of this paper is similar to that first published in this journal by Snygg (1980) \cite{Snygg1980}. However, his theorem uses the Moyal-Wigner formulation and is restricted to Weyl quantisation.  The proof of Theorem I is in the Appendix 3.\

Theorem I: For operator functions ${\hat a}({\hat p},{\hat q})$ of the generic ${\hat p}^r{\hat q}^s{\hat p}^t$ or ${\hat q}^r{\hat p}^s{\hat q}^t$ form such as symmetrised  in accordance to the Born-Jordan scheme eqn(\ref{BJsymmetryrule1}) and eqn(\ref{BJsymmetryrule1a})  or Weyl scheme eqn(\ref{Weylsymmetryrule1a}) and eqn(\ref{Weylsymmetryrule1}) ( see section \ref{Weyl Quantisation}) :

\begin{equation}
Tr\ {\hat a}({\hat p},{\hat q})=\frac{1}{\hbar}\int dp\ dq\ a(p,q)\ .
\label{Theorem}
\end{equation}
Here $a(p,q)$ is a classical function obtained from replacing the phase space operators ${\hat p}$ and ${\hat q}$ in the operator function ${\hat a}({\hat p},{\hat q})$ by their classical variables $p,q$. The integral on the RHS is a classical phase space integral of $a(p,q)$. For our purpose this function is clearly identical with the original classical phase space function $a(p,q)$ as prescribed by the von Neumann rule \cite{vonNeumann1927}.

Using our Theorem I it is straightforward to consider the trace of the LHS of eqn(\ref{Dirac rule1}) so that:
\begin{equation}
 Tr \ [{\hat u },{\hat v}]  =  \frac{1}{\hbar}\int dp\ dq\ \bigl( u(p,q)v(p,q)-v(p,q)u(p,q) \bigr) =0,
\label{Dirac Rule Trace3a}
\end{equation}
as long as the integral is well defined \cite{Starcomment}. Note that the product ${\hat u }{\hat v }$ is not Hermitian, but the commutator is anti-Hermitian (${\hat A}=-{\hat A}^\dag $) i.e. ${\hat A}=i {\hat B}$ where ${\hat B}$ is Hermitian. For further remarks and examples of how this works see Appendix 4.  Next taking the trace of the RHS of eqn(\ref{Dirac rule1}) using Theorem I, we have for the first term:
\begin{equation}
Tr\ \bigl( {\partial \hat u \over \partial \hat q} {\partial \hat v \over \partial \hat p} \bigr) = \frac{1}{\hbar }\int dp\ dq\ ({\partial u \over \partial  q} {\partial  v \over \partial  p} \bigr)\  .
\label{traceP2-1}
\end{equation}
Here both the derivatives and their products are Hermitian (cf eqn(\ref{Hamilton2})), again for an example see Appendix 4.  Integrating by parts w.r.t. to $q$ we have:
\begin{equation}
Tr\ \bigl ({\partial \hat u \over \partial \hat q} {\partial \hat v \over \partial \hat p} \bigr) =\Bigl(\frac{1}{\hbar} \int dp\ u \ {\partial v \over \partial  p}\Bigr)_{\pm L}  -\frac{1}{\hbar}\int dp\ dq\ u \ \bigl({\partial^2  v \over \partial  q \partial  p}\bigr) \  .
\label{traceP2-2}
\end{equation}
For integrable functions that vanish as $u(p,\pm L) \rightarrow 0$ when $L \rightarrow \infty$ the first term vanishes. Also for periodic functions of period $2L$ i.e. $u(p,L)=u(p,-L)$, $v(p,L)=v(p,-L)$ which should also have a Fourier series expansion of the form:
\begin{equation}
 f(p,q) =  \frac{a_0(p)}{2} + \sum_{j=1}^{\infty }\bigl( a_{j}(p) \sin(\frac{j\pi q}{L}) + b_{j}(p) \cos(\frac{j\pi q}{L}) \bigr),
\label{Fourierform1}
\end{equation}
the derivatives w.r.t. to $p$, i.e. $v'(p,L)=v'(p,-L)$ are also periodic and thus the first term also vanishes. At this stage no conditions are required of any function in $p$ space other than that the integral over $p$ must be well defined.

Integrating by parts once more w.r.t. to $p$ for the second term on the RHS eqn(\ref{traceP2-2}), we now have:
\begin{equation}
Tr\ \bigl ({\partial \hat u \over \partial \hat q} {\partial \hat v \over \partial \hat p} \bigr) = -\Bigl(\frac{1}{\hbar} \int dq\ u \ {\partial v \over \partial  q}\Bigr)_{\pm p_F}+\frac{1}{\hbar}\int dp\ dq\ ({\partial v\over \partial  q} {\partial  u \over \partial  p} \bigr).\ \
\label{traceP2-3}
\end{equation}
For integrable functions that vanish as $u(\pm p_F,q) \rightarrow 0$ when $p_F \rightarrow \infty$ the first term vanishes. Now for periodic functions in $p$ space (e.g. harmonic functions) of period $2p_F$  i.e. $u(p_F,q)=u(-p_F,q)$, $v(p_F,q)=v(-p_F,q)$  we need only to consider functions also periodic in $q$ space. For otherwise, the arguments will be the same as in eqn(\ref{Fourierform1}) but with $p\leftrightarrow q$ and $L\leftrightarrow p_F$.  We should now have a Fourier double series expansion of the form:
\begin{equation}
 f(p,q) =   \sum_{k,l=-\infty}^{\infty } a_{k,l}\ \exp(\frac{i\pi k q}{L}) \exp(\frac{i\pi l p}{p_F} ),
\label{Fourierform2}
\end{equation}
with $a_{k,l}=a^*_{-k,-l}$. Then the first term in eqn(\ref{traceP2-3}) vanishes as in eqn(\ref{traceP2-2}). The reader can now see that the remaining term cancels the trace of the second term in eqn(\ref{P2definition}). From these results, it follows that Dirac's quantum Poisson bracket defined in eqn(\ref{P1definition}) is traceless for trace class operators of the type considered here.  It may appear here that the boundary conditions and function types play a more essential role unlike the LHS i.e. eqn(\ref{Dirac Rule Trace3a}), but this is deceptive see further remarks in Appendix 4.  Before concluding this section we should mention that the restriction to the BJ and W Class of operator functions that are consistent with eqn(\ref{Hamilton2}) may perhaps be too stringent since the Dirac quantisation Rule 1 eqn(\ref{Dirac rule1}) is only consistent up to order $O(\hbar)$, see below.  So in general Theorem I could still be used subject to this limitation. This is a subject that requires further investigation.

\section{Other violations}
\label{otherviolations}
In this section we shall look at other Poisson-Lie algebra violations of Dirac's quantisation Rule 1 eqn(\ref{Dirac rule1}). It is easy to show that any quantum Poisson bracket must also satisfy the Leibniz's rule eqn(\ref{LeibnizRule}). This follows from the fact that:
\begin{equation}
\frac{d({\hat F}{\hat G})}{dt}=\frac{d{\hat F}}{dt}{\hat G}+ {\hat F}\frac{d{\hat G}}{dt},
\label{LeibnizRule2}
\end{equation}
maintaining operator ordering in the derivatives and ignoring any explicit time dependencies, without loss of generality. Now from eqn(\ref{Hamilton2a}) we have:
\begin{equation}
\{ {\hat F}{\hat G},{\hat H}\}=\{{\hat F},{\hat H}\}{\hat G} +{\hat F}\{{\hat G},{\hat H}\}.
\label{LeibnizRule3}
\end{equation}
Replacing the Hamiltonian ${\hat H}$ by an arbitrary canonical generator ${\hat W}$,  we have Leibniz's rule:
\begin{equation}
\{ {\hat F}{\hat G},{\hat W}\}=\{{\hat F},{\hat W}\}{\hat G} +{\hat F}\{{\hat G},{\hat W}\},
\label{LeibnizRule4}
\end{equation}
which is a requirement, {\it not} a proof \cite{Kauffman2011}.  It is easy to show that none of the Poisson brackets defined by eqn (\ref{P1definition}), eqn(\ref{P2definition}) or eqn(\ref{P3definition}) satisfy this rule exactly.  For example for $P_1$ we have:
\begin{eqnarray}
\{ {\hat F}{\hat G},{\hat W}\} &=& {\hat F}\{{\hat G},{\hat W}\} + \{{\hat F},{\hat W}\}{\hat G} + [{\hat G},\frac{\partial {\hat W}}{\partial {\hat p}}](\frac{\partial {\hat F}}{\partial {\hat q}}- \frac{\partial {\hat F}}{\partial {\hat p}})\ \nonumber \\
 &=& {\hat F}\{{\hat G},{\hat W}\} + \{{\hat F},{\hat W}\}{\hat G} + i\hbar \{{\hat G},\frac{\partial {\hat W}}{\partial {\hat p}}\}(\frac{\partial {\hat F}}{\partial {\hat q}}- \frac{\partial {\hat F}}{\partial {\hat p}}),\
\label{LeibnizRule5}
\end{eqnarray}
where in the last equation we have replaced the commutator by Rule 1 eqn(\ref{Dirac rule1}).  Substituting the last of eqn(\ref{LeibnizRule5}) into eqn(\ref{Dirac rule1}) again, shows that the violation is of order $O(\hbar^2)$ as in the case of antisymmetry Section \ref{section II}. Again for special cases like ${\hat W}={\hat p}$ or ${\hat W}={\hat q}$, there are no violations as in Section \ref{section II}.  We shall now discuss the Jacobi identity proved in Appendix 1.  Since the essential ingredients for the proof there were bi-linearity and antisymmetry, then it follows that a similar $O(\hbar^2)$ violation ensues for bracket type $P_1$ but {\it not} for bracket types $P_2$ and $P_3$.  We shall not pursue the algebraic significance of these violations in this article. For our purpose, we have shown that Dirac's Rule 1 eqn(\ref{Dirac rule1}) is consistent with a Poisson-Lie algebra , (see Section \ref{Possion-Lie}) eqn(\ref{antisymmetry}) to eqn(\ref{JacobiIdentity}), to order $O(\hbar)$. Without this consistency, Dirac's 1930 derivation \cite{Dirac} of his Rule 1 would have been flawed.

\section{Weyl Quantisation}
\label{Weyl Quantisation}
The discussions of Fourier series in Section \ref{section III} naturally leads one to an examination of the Dirac quantisation scheme for exponential operators: $ {\hat u}({\hat q})= e^{is{\hat q}}$ and $ {\hat v}({\hat p})= e^{it{\hat p}}$ say. Such operators were already studied by Dirac in his early 1926 paper \cite{Dirac4}.  Using Dirac's Rule 1 eqn(\ref{Dirac rule1}) again, we get the following quantisation rule:
\begin{equation}
[e^{is{\hat q}},e^{it{\hat p}}]=i\hbar \{e^{is{\hat q}},e^{it{\hat p}}\}= -i\hbar st e^{is{\hat q}}e^{it{\hat p}} . \
\label{Dirac rule exponential}
\end{equation}
 As this article is devoted to Dirac quantisation, I do not wish to go too deeply into, or indeed be able to do justice, to the role of Weyl quantisation \cite{Weyl1931}. Until recent times this has been the quantisation scheme of choice due to its natural development from the Schr{\"o}dinger representation and von-Neumann's axiomatic foundations \cite{vonNeumann1927}. In this picture we have the well known differential operator ${\hat p}\rightarrow i \hbar \frac{\partial}{\partial q}$ representation for the momentum operator as a realization of the canonical quantisation rule eqn(\ref{BJ Rule}), acting on the Hilbert space of particle wave-functions. Considering quantum kinematics as analogous to an Abelian group of rotations, Weyl \cite{Weyl1931} proposed his quantisation formula in terms of unitary operators ${\hat U}$ and ${\hat V}$ that is nowadays usually written in the form \cite{JordanSudarshan1961}:
\begin{equation}
\ {\hat U}(s){\hat V}(t)\ = e^{-i\hbar s t} \ {\hat V}(t){\hat U}(s). \
\label{Weyl Form}
\end{equation}

Earlier on we briefly mentioned von Neumann's uniqueness theorem \cite{vonNeumannUniqueness} and in particular the failure of bounded operators to form a Heisenberg representation.  It turns out that for unbounded operators ${\hat q}$ and ${\hat p}$, that satisfy the canonical rule eqn(\ref{BJ Rule}),  Weyl's form eqn(\ref{Weyl Form}) for the exponentiated operators ${\hat U(s)}= e^{is{\hat q}}$ , ${\hat V(t)}= e^{it{\hat p}}$ satisfies the Stone-von Neumann uniqueness Theorem, see for example Hall \cite{Hall2013}. This theorem and the Weyl exponentiated form is what makes Weyl quantisation so popular.  The Weyl correspondence formula \cite{Weyl1931,Snygg1980} is an integral transform that allows the mapping from classical to quantum operators and vice-versa. The transform motivates the Moyal-Groeneweld * product algebraic formalism of quantum mechanics which resides entirely in classical phase space. Unfortunately the one-to-one invertibility of this transform while mathematically beautiful to mathematicians, is physically unacceptable to physicists \cite{Kauffman2011,Gosson2014}.  For polynomial operator functions, it leads to a different symmetrisation rule \cite{McCoy1932} from Born and Jordan's eqn(\ref{BJsymmetryrule1}) and eqn(\ref{BJsymmetryrule1a})  namely:
\begin{equation}
{\hat p}^s\ {\hat q}^r \rightarrow  \frac{1}{2^s}\sum_{l=0}^{s} {s\choose l} {\hat p}^{s-l}\ {\hat q}^r\ {\hat p}^l
\label{Weylsymmetryrule1a}
\end{equation}
{\it or}
\begin{equation}
{\hat p}^s\ {\hat q}^r \rightarrow  \frac{1}{2^r}\sum_{j=0}^{r} {r\choose j} {\hat q}^{r-j}\ {\hat p}^s\ {\hat q}^j.
\label{Weylsymmetryrule1}
\end{equation}
Note that both symmetrisation schemes are equivalent for $s+r \leq 2$ but differ when $s \geq 2$ and $r \geq 2$. Currently there are as yet no known experimental conditions to discern between them.   For our purpose we shall first recast the Stone von Neumann version of the Weyl form as a commutator:
\begin{equation}
\ [e^{is{\hat q}},e^{it{\hat p}}]\  = -2i \sin (\frac{\hbar st}{2}) e^{\frac{i\hbar s t}{2}} \ e^{is{\hat q}}\ e^{it{\hat p}} \ .
\label{Stone-VN}
\end{equation}
As can be seen, to the leading order in $\hbar$ this is identical to the Dirac quantisation rule eqn(\ref{Dirac rule exponential}). This can be viewed as an improved quantisation rule that does not have trace anomalies for unbounded operators, following arguments similar to those in Section \ref{section III}.  Unfortunately the scheme is not readily generalisable to spin operators, which are bounded and do not have anomalies. One can see from eqn(\ref{Stone-VN}) that except at order $O(\hbar)$, the commutation rule it represents cannot be derived from any Quantum Poisson bracket rule that we have considered so far.  Moreover the basic quantisation rules eqn(\ref{Weyl Form}) or (\ref{Stone-VN}) are still locked in with the CCR eqn(\ref{BJ Rule}).  This is a prerequisite for the validity of the Weyl correspondence integral transformation formula \cite{Snygg1980,Weyl1931}. The reader will recall that this is fundamentally derived from Dirac's Rule 1, which has inconsistencies at order $\hbar^2$ as shown earlier. Thus the Weyl scheme does not really bring us closer to a fully consistent quantum theory.

An alternative point of view (see the next section) however, is that the original CCR eqn(\ref{BJ Rule}) is considered to be canonical. The improved version eqn(\ref{Stone-VN}) is then considered to be complete as well as being an empirically verified postulate. The extension using Moyal-Groneweld * algebra provides a formalism now known as deformation quantisation which has a natural classical $\hbar \rightarrow 0$ limit to the classical Louville equation in phase space. This is attractive and not so straightforward in the Heisenberg-Born-Jordan-Dirac picture.  Dirac on the other hand was not a great fan of Weyl quantisation nor indeed of the Schr{\"o}dinger picture \cite{Dirac1965}, which he considered to be a ``bad" picture for quantum electrodynamics.  In recent years de Gosson \cite{Gosson2014} has argued that the Schr{\"o}dinger/Heisenberg pictures are only consistent under the Born-Jordan symmetrisation scheme.  He demonstrated that the Born-Jordan scheme provides a resolution of the (nowadays forgotten) angular momentum dilemma and the scheme has possible weak measurement empirical predictions \cite{Gosson2017}. {\it If verified} this would favour Born-Jordan-Dirac instead of Weyl quantisation. The relationship between the trace anomaly discussed in this article with these other issues remains to be further explored.

\section{Conclusion}
\label{Concluding}
Given the results of this study, there are two ways one could see as a resolution of the trace anomaly.  Strategy (A) is to adopt eqn(\ref{Dirac rule1}) as the foundational quantisation rule and adopt an entirely {\it new} algebra that covers both bounded {\it and} unbounded operators. In this algebra the second Dirac rule eqn(\ref{Dirac rule2}) can be eliminated, and more subtle features could emerge from such an algebra, such as in the quantisation of the harmonic oscillator that could be empirically verifiable.  Some current research \cite{Zoran2003} appears to be taking the direction of strategy A including ways to by-pass the Groenewold-van Hove obstruction or no-go theorem of canonical quantisation \cite{Groenewold1946, Hall2013}. This theorem says that a fully consistent canonical quantisation scheme is impossible. A second strategy (B) is to accept that the CCR eqn(\ref{BJ Rule}) should not be read as an equation, but as a (one way) algorithm.  We may all inadvertently have done this for generations by instruction to students: whenever you see a commutator between $\hat q$ and $\hat p$ replace it by $i\hbar$ times a unit operator. This is fine for all practical purposes (FAPP), in the words of John Bell \cite{Bell1990}.

Dirac never gave up on his quantisation scheme. He went on to develop constrained quantisation, using what he called Class I and Class II constraints, and introducing new Dirac brackets \cite{DiracLectures}. These were important advancements, not just in resolving issues in the foundation of quantum electrodynamics but also as a useful tool for curved space-time and gravity \cite{DiracLectures}.  A hint to his mind set can be found in his 1961 article \cite {Dirac1963} in which he proposed the speculation that among the three constants $\hbar,c$ and $e$ of nature, $\hbar$ could someday be replaced but not the other two.   Bell's charge that Dirac was ``perhaps the most distinguished of the `why bother?'s" is in my opinion unjustified.

My thanks to Geoffrey Sewell, Andrew Smith, Anthony Harker and Aris Alexopoulos for assistance, useful comments and careful proof reading. Valuable comments from an anonymous referee of an earlier version of this manuscript is acknowledged.  This paper is dedicated to the memory of a special friend and mentor Professor Wim Caspers of Enschede, the Netherlands (deceased 2010).

\section{Appendix 1: Proof of the Jacobi identity}
\label{Appendix1}
Our proof parallels that of Lemos \cite{Lemos2000}. Consider an infinitesimal canonical transformation $\delta {\hat u}$ generated by an arbitrary generator ${\hat w}$, so that:
\begin{equation}
\delta {\hat u}=\varepsilon \{{\hat u}, {\hat w}\} ,
\label{EqnAppendix2-1}
\end{equation}
where $\varepsilon$ is an infinitesimal parameter. For example $\varepsilon=dt$ and ${\hat w}={\hat H}$ would be an operator contact transformation. The quantum bracket on the RHS of eqn(\ref{EqnAppendix2-1}) can be any one of type $P_1,P_2$ or $P_3$.  Applying the same transformation on a quantum bracket itself, we have:
\begin{equation}
\delta \{{\hat u}, {\hat v}\}=\varepsilon \{\{{\hat u}, {\hat v}\}, {\hat w}\}.
\label{EqnAppendix2-2}
\end{equation}
 Next using bilinearity or the distributive rule of variations we can easily show that:
\begin{equation}
\delta \{{\hat u}, {\hat v}\}= \{\delta{\hat u}, {\hat v}\} + \{{\hat u}, \delta{\hat v}\} ,
\label{EqnAppendix2-3}
\end{equation}
where as usual we have dropped the higher order term: $\{\delta {\hat u}, \delta {\hat v}\}$. Hence from eqn(\ref{EqnAppendix2-1}) we have:
\begin{equation}
\delta \{{\hat u}, {\hat v} \}= \{ \varepsilon \{ {\hat u}, {\hat w}\} , {\hat v} \} + \{ {\hat u}, \varepsilon \{{\hat v}, {\hat w}\} \} \} .
\label{EqnAppendix2-4}
\end{equation}
Equating equations (\ref{EqnAppendix2-2}) and (\ref{EqnAppendix2-4}) we have:
\begin{equation}
 \{\{{\hat u}, {\hat v}\}, {\hat w}\}= \{ \{ {\hat u}, {\hat w}\} , {\hat v} \} + \{ {\hat u}, \{{\hat v}, {\hat w}\} \} \} .
\label{EqnAppendix2-5}
\end{equation}
Finally using antisymmetry we have the Jacobi identity:
\begin{equation}
\{\{u,v\},w\} + \{\{v,w\},u\}  + \{\{w,u\},v\}  =0.
\label{EqnAppendix2-6}
\end{equation}
Note that because antisymmetry is the key ingredient for the proof, violations to the Jacobi identity are also of order $\hbar^2$ for bracket type $P_1$.  There are no violations for bracket types $P_2$ and $P_3$.
\section{Appendix 2: Proof that the BJ and W class functions satisfy the canonical equations (\ref{Hamilton2}) without using the CCR eqn(\ref{BJ Rule}). }\
\label{Appendix2}
The Born-Jordan symmetrisation rule eqn(\ref{BJsymmetryrule1}) was originally proved in their seminal paper using the canonical quantisation rule eqn(\ref{BJ Rule}) and their extensions for polynomial operators. Let us first review this proof.  By induction from eqn(\ref{BJ Rule}), they first show that:
\begin{equation}
 [{\hat q}^n, {\hat p}^m]= i \hbar m \sum_{l=0}^{n-1} {\hat q}^{n-1-l}{\hat p}^{m-1}{\hat q}^l .
\label{EqnAppendix3-1}
\end{equation}
From eqn(\ref{BJ Rule}), they show by interchanging ${\hat q}\leftrightarrow {\hat p}$ and the sign of $\hbar$, that this is also equivalent to:
\begin{equation}
 [{\hat q}^n, {\hat p}^m]= i \hbar n \sum_{j=0}^{m-1} {\hat p}^{m-1-j}{\hat q}^{n-1}{\hat p}^j .
\label{EqnAppendix3-2}
\end{equation}
Equating eqn(\ref{EqnAppendix3-1}) and eqn(\ref{EqnAppendix3-2}) gives the useful relation:
\begin{equation}
 \frac{1}{r+1} \sum_{j=0}^{r} {\hat q}^{r-j}{\hat p}^s {\hat q}^j= \frac{1}{s+1} \sum_{l=0}^s {\hat p}^{s-l}{\hat q}^r{\hat p}^l .
\label{EqnAppendix3-3}
\end{equation}
Then they constructed the Hamiltonian function:
\begin{equation}
{\hat H} = \frac{1}{s+1} \sum_{l=0}^s {\hat p}^{s-l}{\hat q}^r{\hat p}^l=\frac{1}{r+1} \sum_{j=0}^{r} {\hat q}^{r-j}{\hat p}^s {\hat q}^j,
\label{EqnAppendix3-4}
\end{equation}
by eqn(\ref{EqnAppendix3-3}).  Now they first show that this satisfies the second of eqn(\ref{Hamilton2}) namely:
\begin{equation}
  [{\hat p},{\hat H} ]= - {i\hbar} {\partial \hat H \over \partial \hat q}.
\label{EqnAppendix3-5}
\end{equation}
To do this they substitute the first of eqn(\ref{EqnAppendix3-4}) into the commutator on the LHS of eqn(\ref{EqnAppendix3-5}) which gives:
 \begin{equation}
  [{\hat p},{\hat H} ]= \frac{1}{s+1} [{\hat p}^{s+1},{\hat q}^r ]=-i \hbar  \sum_{l=0}^{r-1} {\hat q}^{r-1-l}{\hat p}^{s}{\hat q}^l,
\label{EqnAppendix3-6}
\end{equation}
where the last equation follows from eqn(\ref{EqnAppendix3-1}). Then by the algorithm for Born-Jordan matrix differentiation \cite {BornJordan1925} on ${\hat H}$ (see later) they show that this is equivalent to the RHS of eqn(\ref{EqnAppendix3-5}). Next again by evaluating the commutator for the first of eqn(\ref{Hamilton2})  using now the second of eqn(\ref{EqnAppendix3-4}), they show that:
\begin{equation}
[{\hat q},{\hat H} ]= \frac{1}{r+1} [{\hat q}^{r+1},{\hat p}^s ]=i \hbar \sum_{j=0}^{s-1} {\hat p}^{s-1-j}{\hat q}^{r}{\hat p}^j.
\label{EqnAppendix3-7}
\end{equation}
where the last equation follows from eqn(\ref{EqnAppendix3-2}). Once again by using the Born-Jordan algorithm for matrix differentiation \cite {BornJordan1925} ( see below) on ${\hat H}$ they show that this is equivalent to:
\begin{equation}
[{\hat q},{\hat H} ]=  {i\hbar} {\partial \hat H \over \partial \hat p},
\label{EqnAppendix3-8}
\end{equation}
thus completing their proof.

For our purpose we must also prove eqn(\ref{EqnAppendix3-1}), eqn(\ref{EqnAppendix3-2}) and the equality of eqn(\ref{EqnAppendix3-4}) but we do not have the luxury of using the trick of changing from eqn(\ref{EqnAppendix3-1}) to eqn(\ref{EqnAppendix3-2}) above which was based on the CCR eqn(\ref{BJ Rule}). However we are at liberty to swap ${\hat q}\leftrightarrow {\hat p}$ in the BJ symmetrisation rule eqn(\ref{BJsymmetryrule1}) so that:
\begin{equation}
{\hat q}^s {\hat p}^r \rightarrow \frac{1}{s+1}\sum_{l=0}^{s}{\hat q}^{s-l}\ {\hat p}^r\ {\hat q}^l.
\label{BJsymmetryrule2}
\end{equation}
Now we must first evaluate the commutator on the LHS of eqn(\ref{EqnAppendix3-2}) using Dirac's Rule 1 eqn(\ref{Dirac rule1}), which gives:
\begin{equation}
 [{\hat q}^n, {\hat p}^m]= i\hbar \{{\hat q}^n,{\hat p}^m\}  = i\hbar n m {\hat q}^{n-1} {\hat p}^{m-1}.
\label{EqnAppendix3-9}
\end{equation}
This result must be symmetrised. On applying the symmetrisation rule eqn(\ref{BJsymmetryrule2}), eqn(\ref{EqnAppendix3-1}) is proved.
Similarly we can obtain:
\begin{equation}
 [{\hat p}^m, {\hat q}^n]= i\hbar \{{\hat p}^m,{\hat q}^n\}  = -i\hbar m n {\hat p}^{m-1} {\hat q}^{n-1}.
\label{EqnAppendix3-10}
\end{equation}
Here we apply the symmetrisation rule eqn(\ref{BJsymmetryrule1}) and eqn(\ref{EqnAppendix3-2}) is also proved.   Equating eqn(\ref{EqnAppendix3-9}) with eqn(\ref{EqnAppendix3-10})  now proves the equality eqn(\ref{EqnAppendix3-4}). The rest of the proof follows as before unchanged, thus proving that the Born-Jordan symmetrisation rule eqn(\ref{BJsymmetryrule1}) satisfies the canonical equations eqn(\ref{Hamilton2}), without the CCR eqn(\ref{BJ Rule}).  By generalising to an arbitrary canonical generator ${\hat u},{\hat v}$ in place of the above Hamiltonian $H$ the results eqn(\ref{Hamilton2}) are proved. Following the same arguments, the equality of eqn(\ref{Weylsymmetryrule1a}) and eqn(\ref{Weylsymmetryrule1}) for Weyl quantisation can also be proved without using the CCR eqn(\ref{BJ Rule}). The latter was a requirement in McCoy's derivation \cite{McCoy1932}.

The above proof follows the historical developments of Born and Jordan 1925 \cite{BornJordan1925}. A more direct proof of the above results can also be obtained using Heisenberg's simpler definition of the matrix differentiation operator \cite{Born-Jordan-Heisenberg1926}, which we shall not repeat here. However the reader must note that this only works because Heisenberg's derivative and Born-Jordan's derivative are identical \cite{Gosson2016}.  The latter is given by the following algorithm: (i) first all products must be written out in full, e.g. ${\hat p^3}{\hat q^2}={\hat p}{\hat p}{\hat p}{\hat q}{\hat q}$ (ii) for the derivative of each member, delete all products prior to it, but keep them in storage whilst maintaining all orderings (iii) append the items in storage in front of the products of the chosen member in (ii).  Sum and repeat (i) to (iii) until all members are differentiated. Finally the result can be condensed by restoring all exponents. For example for ${\hat y}={\hat x}_1^2{\hat x}_2{\hat x}_1{\hat x}_3$ the derivative $\frac{\partial {\hat y}}{\partial {\hat x}_1}={\hat x}_1{\hat x}_2x_1{\hat x}_3+{\hat x}_2{\hat x}_1{\hat x}_3{\hat x}_1+{\hat x}_3{\hat x}_1^2{\hat x}_2$ etc. Heisenberg's definition of the derivative is given by \cite{Born-Jordan-Heisenberg1926}:
\begin{equation}
\frac{\partial {\hat f}}{\partial {\hat x}_i} = \lim_{\alpha\rightarrow 0}\frac{{\hat f}({\hat x}_1,{\hat x}_2,.,{\hat x}_i+\alpha {\hat I},..,{\hat x}_n)-{\hat f}({\hat x}_1,{\hat x}_2,...,{\hat x}_n)}{\alpha}.
\label{EqnAppendix3-10c}
\end{equation}
By employing this definition, an elegant proof of the validity of eqn(\ref{Hamilton2}) for Weyl symmetrisation can be obtained. We shall employ this definition of Heisenberg's derivative to prove eqn(\ref{Hamilton2}) as an example. Using the Weyl symmetrisation rule eqn(\ref{Weylsymmetryrule1}) we can evaluate the formula for the commutator in eqn(\ref{EqnAppendix3-2}) as:
\begin{eqnarray}
 [{\hat p}, {\hat H}]&=& \frac{1}{2^s}\sum_{l=0}^{s} {s\choose l} {\hat p}^{s-l}\ [{\hat p},{\hat q}^r]\ {\hat p}^l  =\frac{i\hbar }{2^s}\sum_{l=0}^{s} {s\choose l}
 {\hat p}^{s-l}\ \{{\hat p},{\hat q}^r \} \ {\hat p}^l ,\ \nonumber \\
 &=& \frac{-i\hbar }{2^s}\ r \sum_{l=0}^{s} {s\choose l} {\hat p}^{s-l}\ {\hat q}^{r-1}\ {\hat p}^l  = {-i\hbar} {\partial \hat H \over \partial \hat q}.
\label{EqnAppendix3-10a}
\end{eqnarray}
The first equality follows from commutator algebra. The second equality follows from the Dirac's Rule 1 and the last equality from Heisenberg's definition \cite{Born-Jordan-Heisenberg1926} of the matrix differential eqn(\ref{EqnAppendix3-10c}) acting on eqn(\ref{Weylsymmetryrule1a}). Alternatively Dirac's Rule 1 can be applied from the start and Leibniz's rule invoked (see section \ref{otherviolations}).  The proof is not yet complete.  We must also show that for consistency the last equality is identical with Heisenberg's derivative eqn(\ref{EqnAppendix3-10c}) acting on eqn(\ref{Weylsymmetryrule1}). That this is the case can be accomplished with some straightforward algebra involving the combinatorial factors.   In a similar way eqn(\ref{EqnAppendix3-8}) is proved for Weyl symmetrisation for Heisenberg's differentiation. The above arguments have shown that commutator bracket algebra plus Dirac's Rule 1 is fully compatible with Heisenberg's differentiation eqn(\ref{EqnAppendix3-10c}) for Weyl symmetrisation. Also we know that Heisenberg's derivative and Born-Jordan's derivative are {\it not} identical for Weyl symmetrisation\cite{Gosson2016}.  In fact it is easy to show that Heisenberg's derivative for Weyl symmetrisation is given by:
\begin{equation}
\frac{\partial}{\partial {\hat p} } \Bigl( \frac{1}{2^s}\sum_{l=0}^{s} {s\choose l} {\hat p}^{s-l}\ {\hat q}^r\ {\hat p}^l \Bigr ) = \frac{s}{2^{s-1}}\sum_{l=0}^{s-1} {s-1\choose l} {\hat p}^{s-1-l}\ {\hat q}^r\ {\hat p}^l= \frac{s}{2^r}\sum_{j=0}^{r} {r\choose j} {\hat q}^{r-j}\ {\hat p}^{s-1}\ {\hat q}^j.
\label{EqnAppendix3-11a}
\end{equation}
Whereas using the algorithm above, Born-Jordan's derivative for Weyl symmetrisation is given by the unique formula:
\begin{equation}
\frac{\partial}{\partial {\hat p} } \Bigl( \frac{1}{2^s}\sum_{l=0}^{s} {s\choose l} {\hat p}^{s-l}\ {\hat q}^r\ {\hat p}^l \Bigr )   = \sum_{l=0}^{s-1} {\hat p}^{s-1-l}\ {\hat q}^r\ {\hat p}^l.
\label{EqnAppendix3-11b}
\end{equation}
Therefore we can conclude without further work, that Weyl symmetrisation does {\it not} satisfy the Hamilton-like equations eqn(\ref{Hamilton2}) for Born-Jordan's derivatives when $s\geq 3$ and $r\geq 2$ . However it is easy to show, see Appendix 4, that the discrepancy between these derivatives is again of a higher order in $\hbar$ e.g. for $s=3,r=2$ we have a discrepancy of $\frac{1}{2} \hbar^2$ and for $s=3,r=3$, we have a discrepancy of $\frac{3}{2} \hbar^2 {\hat q}$. Given that there are already so many violations of order $\hbar^2$, we shall not treat Weyl symmetrisation differently in this paper, unless explicitly required.

Next it is easy to show that if ${\hat u},{\hat v}$ are of the BJ or W class, then their bi-products e.g. ${\hat u}{\hat v}$, and bi-products of their (compatible) derivatives e.g. ${\partial \hat u \over \partial \hat p}{\partial \hat v \over \partial \hat q} $ and ${\partial \hat u \over \partial \hat q}{\partial \hat v \over \partial \hat p} $  are also of this class.  This means that these products are all ultimately expressible (albeit with much tedious work), as linear sums of the generic ${\hat p}^s{\hat q}^r{\hat p}^t$ or ${\hat q}^s{\hat p}^r{\hat q}^t$ type as required for the application of Theorem I in section \ref{section III}.  The reader can easily prove this for the product ${\hat u}{\hat v}$. First we take the derivative:
\begin{eqnarray}
\frac{\partial}{\partial {\hat p}} ({\hat u}{\hat v})&=& \frac{\partial {\hat u}}{\partial {\hat p}} {\hat v}+{\hat u}\frac{\partial {\hat v}}{\partial {\hat p}}\ \nonumber \\
 &=& \frac{1}{i\hbar}[{\hat q},{\hat u}]{\hat v}+\frac{1}{i\hbar}{\hat u}[{\hat q},{\hat v}]= \frac{1}{i\hbar}[{\hat q},{\hat u}{\hat v}].
\label{EqnAppendix3-11}
\end{eqnarray}
The first equation follows from the rules of matrix differentiation \cite{BornJordan1925,Dirac3}, the second follows from the definition of the BJ and W Class and the last follows from the algebra of commutators. Similarly we can prove that:
\begin{eqnarray}
\frac{\partial}{\partial {\hat q}} ({\hat u}{\hat v})&=& \frac{\partial {\hat u}}{\partial {\hat q}} {\hat v}+{\hat u}\frac{\partial {\hat v}}{\partial {\hat q}}\ \nonumber \\
 &=& -\frac{1}{i\hbar}[{\hat p},{\hat u}]{\hat v}-\frac{1}{i\hbar}{\hat u}[{\hat p},{\hat v}]= -\frac{1}{i\hbar}[{\hat p},{\hat u}{\hat v}].
\label{EqnAppendix3-12}
\end{eqnarray}
Thus we have proved that the product ${\hat u}{\hat v}$ is also of the BJ and W Class. This statement requires some qualification for as stated in the text ${\hat u}{\hat v}$  is not Hermitian.  However any such operator can be written as a sum of a Hermitian and an anti-Hermitian operator i.e. ${\hat u}{\hat v}={\hat A}+{\hat C}={\hat A}+i{\hat B}$ where ${\hat A}$ and ${\hat B}$ are Hermitian and ${\hat C}$ is anti-Hermitian. Thus ${\hat A}$ and ${\hat B}$ are both of the BJ and W Class.    That the same property holds for the bi-product of their derivatives follows by the same manipulations.  For the avoidance of doubt, we shall just show this for the product: ${\hat w}={\partial \hat u \over \partial \hat p}{\partial \hat v \over \partial \hat q}$. Taking the first derivative w.r.t. to $\hat p $ we have:
\begin{eqnarray}
\frac{\partial {\hat w}}{\partial {\hat p}}&=& \frac{\partial^2 {\hat u}}{\partial {\hat p}^2} \frac{\partial {\hat v}}{\partial {\hat q}}+ \frac{\partial {\hat u}}{\partial {\hat p}}\frac{\partial^2 {\hat v}}{\partial {\hat p}\partial {\hat q}}\ \nonumber \\
&=&\frac{1}{i\hbar}[{\hat q},\frac{\partial {\hat u}}{\partial {\hat p}}]\frac{\partial {\hat v}}{\partial {\hat q}}+\frac{1}{i\hbar}\frac{\partial {\hat u}}{\partial {\hat p}}[{\hat q},\frac{\partial {\hat v}}{\partial {\hat q}}] \ \nonumber \\
 &=&\frac{1}{i\hbar }[{\hat q},{\hat w}].
\label{EqnAppendix3-13}
\end{eqnarray}
Similarly taking the first derivative wrt to $\hat q $, we have:
\begin{eqnarray}
\frac{\partial {\hat w}}{\partial {\hat q}}&=& \frac{\partial^2 {\hat u}}{\partial {\hat q}\partial {\hat p}} \frac{\partial {\hat v}}{\partial {\hat q}}+ \frac{\partial {\hat u}}{\partial {\hat p}}\frac{\partial^2 {\hat v}}{\partial {\hat q}^2}\ \nonumber \\
&=&-\frac{1}{i\hbar}[{\hat p},\frac{\partial {\hat u}}{\partial {\hat p}}]\frac{\partial {\hat v}}{\partial {\hat q}}-\frac{1}{i\hbar}\frac{\partial {\hat u}}{\partial {\hat p}}[{\hat p},\frac{\partial {\hat v}}{\partial {\hat q}}] \ \nonumber \\
 &=&-\frac{1}{i\hbar }[{\hat p},{\hat w}].
\label{EqnAppendix3-14}
\end{eqnarray}
The reader will have no difficulty proving the same for ${\hat z}={\partial \hat u \over \partial \hat q}{\partial \hat v \over \partial \hat p}$.  It is easy to show from eqn(\ref{Hamilton2}) that all the derivatives and the product of the derivatives considered here are Hermitian . Appendix 4 contains some examples to illustrate how all these results work in practice.

\section{Appendix 3: Proof of Theorem I}
\label{Appendix3}
The proof of the Theorem in the text namely eqn (\ref{Theorem}) begins with:
\begin{equation}
Tr\ {\hat a}({\hat p},{\hat q})=\int dq\ dq'\ <q'|{\hat a}({\hat p},{\hat q})|q>\delta(q-q').\
\label{AppendixEqn1}
\end{equation}
For any arbitrary quantum operator function ${\hat a}({\hat p},{\hat q})$ expressed as a polynomial or power series, one can always reorder all operators so that all ${\hat q}$'s are on the right and all ${\hat p}$'s are on the left or vice versa using commutation relations of the type eqn(\ref{BJ Rule}) or eqn(\ref{Dirac rule1}), or relations derived from them in the text. This way one gets a series of monomial terms with successively lower degrees in the operators but higher degrees in $\hbar$.  For example, using eqn(\ref{BJ Rule}) we can easily obtain: ${\hat q}{\hat p}{\hat q}={\hat p}{\hat q}^2+ i\hbar {\hat q}$ or ${\hat p}^2{\hat q}{\hat p}{\hat q}{\hat p}={\hat p}^4{\hat q}^2+ 3i\hbar {\hat p}^3{\hat q}-\hbar^2{\hat p}^2$ and so on.  We will assume that for an infinite series, such a process can also be made, so that the following theorem can be applied so long as we keep track of our $\hbar$ terms. Therefore in principle the theorem below goes unchanged, except that at the end of the day, the function ${\tilde a}(p,q)$ will {\it not} be the same as the von-Neumann postulated classical object $a(p,q)$. An algorithm can be developed for this \cite{Snygg1980} but already we see there are problems. For the above example ${\hat q}{\hat p}{\hat q}$ we can easily see that we can have two answers.  As Hermitian operators can act both on the left and on the right in the trace, no re-ordering is really necessary.  However if we do re-order we pick up an extra $i\hbar {\hat q}$ term.  Are such inconsistencies a logical or mathematical flaw in quantum mechanics?  Almost a century ago debates such as this were very heated. Mathematicians \cite{Temple1935} were upset, while eminent physicists such as Peierls \cite{Peierls1935} suggested that there is no logical flaw. Instead the different results can be identified with different ``apparatus" or experimental conditions \cite{Peierls1935}.  I shall leave the reader to contemplate on this Copenhagen interpretation for another day.    For our purpose as described in the text, we have deliberately chosen functions of the BJ and W Class types in which such operator re-orderings are already imposed a prior. To avoid misunderstandings, we shall denote such a function type as: ${\hat a}({\hat q},{\hat p},{\hat q})$ for ${\hat q}^s{\hat p}^r{\hat q}^t$ forms such as eqn(\ref{BJsymmetryrule1}), eqn(\ref{Weylsymmetryrule1}) and ${\hat a}({\hat p},{\hat q},{\hat p})$ for ${\hat p}^s{\hat q}^r{\hat p}^t$ forms such as eqn(\ref{BJsymmetryrule1a}), eqn(\ref{Weylsymmetryrule1a}). Let us consider the first form.

Now the action of the operator ${\hat a}$ on the state $|q>$ to the right will give ${\hat a}({\hat q},{\hat p},{\hat q})|q>={\hat a}({\hat q},{\hat p},q)|q>$ and on the state $<q|$ to the left will give $<q|{\hat a}({\hat q},{\hat p},{\hat q})=<q|{\hat a}(q,{\hat p},{\hat q})$ since the states $|q>$ are eigenvectors of ${\hat q}$. Thus we have:
\begin{equation}
Tr\ {\hat a}({\hat p},{\hat q})=\int dq\ dq'\ <q'|{\hat a}(q',{\hat p},q)|q>\delta(q-q').\
\label{AppendixEqn2}
\end{equation}
We can now introduce a complete set of momentum states $|p>$ where $<p|q>=\frac{1}{\sqrt{\hbar}}\ e^{i\frac{pq}{\hbar}}$ and insert the identity operator: $\int dp\ |p><p|$ into eqn(\ref{AppendixEqn2}) as before so that:
\begin{eqnarray}
Tr\ {\hat a}&=& \int dq\ dq'\ dp \ <q'|{\hat a}(q',{\hat p},q)|p><p|q>\delta(q-q')\\
    & = &  \int  dq\ dq'\ \frac{dp}{\sqrt{\hbar}}\ <q'|{\hat a}(q',{\hat p},q))|p>\ e^{i\frac{pq}{\hbar}}\delta(q-q')\\
    & = & \int dq\ dq'\ \frac{dp}{\hbar} a(q',p,q)\ e^{i \frac{p(q-q')}{\hbar}}\delta(q-q').
\label{AppendixEqn3}
\end{eqnarray}
Integration over $q'$ is now trivial, provided the $p$ integration does not diverge, and we finally have:
\begin{equation}
Tr\ {\hat a}({\hat p},{\hat q})=\frac{1}{\hbar}\int dp\ dq\ a(p,q)\,
\label{AppendixEqn4}
\end{equation}
since $a(q,p,q)=a(p,q)$.  For the second form of eqn(\ref{BJsymmetryrule1a}) or eqn(\ref{Weylsymmetryrule1a})  i.e. for the ${\hat p}^s{\hat q}^r{\hat p}^t$ type operators we can interchange momentum and position representations in the above argument, performing the trace first in momentum space in eqn(\ref{AppendixEqn2}) instead and the same arguments follow. This completes the proof of the theorem in the text for the BJ and W Class of functions.

\section{Appendix 4: Some Examples}
\label{Appendix4}
In the first example we shall consider two non-separable functions ${\hat u}=\frac{1}{2}({\hat p}{\hat q}+{\hat q}{\hat p})$ and ${\hat v}=\frac{1}{2}({\hat p}^2{\hat q}+{\hat q}{\hat p}^2)$ to illustrate how the results of Appendix 2 work in practice.
First we shall consider Born-Jordan symmetrisation. Here we have two formulas for ${\hat v}=\frac{1}{2}({\hat p}^2{\hat q}+{\hat q}{\hat p}^2)$ or ${\hat v}=\frac{1}{3}({\hat p}^2{\hat q}+{\hat p}{\hat q}{\hat p}+{\hat q}{\hat p}^2)$ .  Let us use the first formula, then the products: ${\hat u}{\hat v}$ and ${\hat v}{\hat u}$ are now given by:
\begin{eqnarray}
{\hat u}{\hat v} &=& \frac{1}{4}({\hat p}{\hat q}{\hat p}^2 {\hat q}+{\hat p}{\hat q}^2{\hat p}^2+{\hat q}{\hat p}^3{\hat q}+{\hat q}{\hat p}{\hat q}{\hat p}^2) \nonumber \\
&=& \frac{1}{2} ({\hat q}{\hat p}^3 {\hat q} + {\hat p}{\hat q}^2 {\hat p}^2)-\frac{\hbar^2}{2}{\hat p}  \\
{\hat v}{\hat u} &=& \frac{1}{4}({\hat p}^2{\hat q}{\hat p}{\hat q} +{\hat p}^2{\hat q}^2{\hat p} +{\hat q}{\hat p}^3{\hat q}+{\hat q}{\hat p}^2{\hat q}{\hat p}) \nonumber \\
&=& \frac{1}{2} ({\hat q}{\hat p}^3 {\hat q} + {\hat p}^2{\hat q}^2 {\hat p})-\frac{\hbar^2}{2}{\hat p}.
\label{Appendix4-1}
\end{eqnarray}
Here we have used the CCR eqn(\ref{BJ Rule}) or Dirac's Rule 1 eqn(\ref{Dirac rule1}) which are equivalent in this case for the reduction of the four operator terms.  Now the Hermitian and $\hbar^2$ terms cancel and we have for the commutator:
\begin{equation}
[{\hat u},{\hat v}]= \frac{1}{2} ({\hat p}{\hat q}^2 {\hat p}^2 -{\hat p}^2{\hat q}^2 {\hat p}).
\label{Appendix4-2}
\end{equation}
It is now straightforward to evaluate the derivatives and take their products, which must be symmetrised to be Hermitian:
\begin{eqnarray}
\frac {\partial {\hat u}}{\partial {\hat q}}\frac {\partial {\hat v}}{\partial {\hat p}}  &=&  \frac{1}{2} ({\hat p}^2{\hat q}+{\hat q}{\hat p}^2)+ {\hat p}{\hat q}{\hat p}\ \nonumber \\
\frac {\partial {\hat u}}{\partial {\hat p}}\frac {\partial {\hat v}}{\partial {\hat q}}  &=&  \frac{1}{2} ({\hat p}^2{\hat q}+{\hat q}{\hat p}^2).
\label{Appendix4-3}
\end{eqnarray}
Thus the difference of the derivatives is given by:
\begin{equation}
\frac {\partial {\hat u}}{\partial {\hat q}}\frac {\partial {\hat v}}{\partial {\hat p}} - \frac {\partial {\hat u}}{\partial {\hat p}}\frac {\partial {\hat v}}{\partial {\hat q}} = {\hat p}{\hat q}{\hat p}.
\label{Appendix4-4}
\end{equation}
A further reduction of eqn(\ref{Appendix4-2}) gives:
\begin{equation}
[{\hat u},{\hat v}]= \frac{1}{2} ({\hat p}{\hat q}^2 {\hat p}^2 -{\hat p}^2{\hat q}^2 {\hat p})= \frac{1}{2} ({\hat p}[{\hat q}^2, {\hat p} ] {\hat p}) =i\hbar {\hat p}{\hat q}{\hat p} ,
\label{Appendix4-5}
\end{equation}
showing the consistency of Born-Jordan's symmetrisation with Dirac's Rule 1. We shall not repeat this exercise with the second formula ${\hat v}=\frac{1}{3}({\hat p}^2{\hat q}+{\hat p}{\hat q}{\hat p}+{\hat q}{\hat p}^2)$ but merely quote the final result:
\begin{equation}
[{\hat u},{\hat v}]= \frac{1}{3} ({\hat p}{\hat q}^2 {\hat p}^2 -{\hat p}^2{\hat q}^2 {\hat p})+ \frac{i\hbar}{6}({\hat p}^2{\hat q} +{{\hat q}\hat p}^2)=i\hbar {\hat p}{\hat q}{\hat p}.
\label{Appendix4-6}
\end{equation}

Now for Weyl's symmetrisation we have two formulas for $v$: ${\hat v}=\frac{1}{2}({\hat p}^2{\hat q}+{\hat q}{\hat p}^2)$ or ${\hat v} =\frac{1}{4}({\hat p}^2{\hat q}+2{\hat p}{\hat q}{\hat p} +{\hat q}{\hat p}^2)$. For the first formula the calculation is identical with the Born-Jordan results given above, since all derivatives are identical for $s=2$ chosen here.  We only need to perform the calculation for the second formula which gives:
\begin{eqnarray}
{\hat u}{\hat v} &=& \frac{1}{8}({\hat p}{\hat q}{\hat p}^2 {\hat q}+2{\hat p}{\hat q}{\hat p}{\hat q}{\hat p}+{\hat p}{\hat q}^2{\hat p}^2+{\hat q}{\hat p}^3 {\hat q}+2{\hat q}{\hat p}^2{\hat q}{\hat p} +{\hat q}{\hat p}{\hat q}{\hat p}^2)     \\
{\hat v}{\hat u} &=& \frac{1}{8}({\hat p}^2{\hat q}{\hat p} {\hat q}+2{\hat p}{\hat q}{\hat p}{\hat q}{\hat p}+{\hat p}^2{\hat q}^2{\hat p}+{\hat q}{\hat p}^3 {\hat q}+2{\hat p}{\hat q}{\hat p}^2{\hat q} +{\hat q}{\hat p}^2{\hat q}{\hat p}).
\label{Appendix4-7}
\end{eqnarray}
Finally the Hermitian terms will always cancel and a reduction using the CCR eqn(\ref{BJ Rule}) gives the final result for the commutator as:
\begin{equation}
[{\hat u},{\hat v}]= \frac{1}{4} ({\hat p}{\hat q}^2 {\hat p}^2 -{\hat p}^2{\hat q}^2 {\hat p})+ \frac{i\hbar}{4}({\hat p}^2{\hat q} +{{\hat q}\hat p}^2)=i\hbar {\hat p}{\hat q}{\hat p}.
\label{Appendix4-8}
\end{equation}
It is straight forward to evaluate the derivatives and the results are identical with equations (\ref{Appendix4-3}) and (\ref{Appendix4-4}) showing the consistency of Weyl's symmetrisation with Dirac's Rule 1. Note that the answer for the commutators can be expressed in several ways. First as an anti-Hermitian operator such as eqn(\ref{Appendix4-2}) which does not manifest $\hbar$ explicitly or secondly as $i\hbar$ times a Hermitian operator such as eqn(\ref{Appendix4-5}) which does manifest $\hbar$ explicitly or thirdly as a combination of both such as eqn(\ref{Appendix4-6}).  In the first case the trace will lead to zero by Theorem I while in the second case it will lead to an integral similar to eqn(\ref{traceP2-1}) which will require boundary considerations. There is no contadiction here as convergence issues are not treated in this paper. These comments will clarify remarks after eqn(\ref{Dirac Rule Trace3a}) in the text.  The reader is invited to repeat this exercise for the function ${\hat v}=\frac{1}{4}({\hat q}^2{\hat p}^3+2{\hat q}{\hat p}^3{\hat q}+{\hat p}^3{\hat q}^2)=\frac{1}{8}({\hat p}^3{\hat q}^2+3{\hat p}^2{\hat q}^2{\hat p}+3{\hat p}{\hat q}^2{\hat p}^2 +{\hat q}^2{\hat p}^3)$ which are equivalent under Weyl's symmetrisation. Here the Born-Jordan and Weyl derivatives are no longer identical. It is easily verified that they are given by:
\begin{eqnarray}
(\frac {\partial {\hat v}}{\partial {\hat p}})_{\rm Heis}  &=&  \frac{3}{4} ({\hat p}^2{\hat q}^2+ 2{\hat p}{\hat q}^2{\hat p}+{\hat q}^2{\hat p}^2)=\frac{3}{2} ({\hat p}^2{\hat q}^2+{\hat q}^2{\hat p}^2)+\frac{3}{2}\hbar^2\ \nonumber \\
(\frac {\partial {\hat v}}{\partial {\hat p}})_{\rm BJ}  &=&  {\hat p}^2{\hat q}^2+ {\hat p}{\hat q}^2{\hat p}+{\hat q}^2{\hat p}^2=\frac{3}{2} ({\hat p}^2{\hat q}^2+{\hat q}^2{\hat p}^2)+\hbar^2.
\label{Appendix4-9}
\end{eqnarray}
As a consequence, this difference will result in a discrepancy in eqn(\ref{Appendix4-4}) of $\frac{1}{2} \hbar^2 {\hat p}$, if we use the Born-Jordan derivatives.


\begin{thebibliography}{10}
\bibitem{endnote1} In this paper all quantities with top hats are quantum operators, those without are classical variables and for ease of notation we shall consider a one-dimensional system of length $L$. Generalisation to higher dimensions can be made by adding subscript indices for Cartesian components.
\bibitem{endnote2} A Poisson algebra is an associative algebra defined on a Poisson manifold that is a special case of a Lie algebra with a Lie bracket given by the Poisson bracket. Historically Poisson algebra came first but we shall use the name Poisson-Lie defined here by the algebra equations (\ref{antisymmetry} to \ref{JacobiIdentity}). Dirac was the first to seek an extension of this algebra to a manifold of quantum operators defined over a Hilbert space.
\bibitem{traceclassI} For Hermitian operators $u,v$ which Dirac considered, trace class simply implies boundedness: $Tr\ |u| < \infty, Tr\ |v| < \infty$, where $|u|$,$|v|$ denote the norms .
\bibitem{endnote4} The consideration of periodic functions will require extensions to more sophisticated double series such as a Fourier or Bessel double series. However this will require more powerful techniques than what we have presented here. As such we shall assume that all periodic functions encountered (such as $e^{ikq}$) must be treated as an infinite power series.
\end{thebibliography}
\end{document}